# Using an Instrumented Hammer to predict the lateral hinge fracture in high tibial osteotomy: a cadaveric study

Claire Bastard[1], Manon Bas dit Nugues[2], Giuseppe Rosi[3], Charles Henri Flouzat-Lachaniette[4], Arnaud Dubory[4] and Guillaume Haiat[2,*]

[1] Sorbonne Université, Paris, France. Service de Chirurgie de la Main, Service d'Orthopédie et de Traumatologique, Hôpital Saint-Antoine, Sorbonne Université, 184, rue du Faubourg-Saint-Antoine, 75012 Paris, France.

[2]CNRS, Univ Paris Est Creteil, Univ Gustave Eiffel, UMR 8208, MSME, 61, Avenue du Général de Gaulle, F-94010 Créteil Cedex, France

[3]Univ Paris Est Creteil, Univ Gustave Eiffel, CNRS, UMR 8208, MSME, F-94010 Créteil, France

[4] INSERM U955, IMRB Université Paris-Est, 51 avenue du Maréchal de Lattre de Tassigny, 94000 Créteil, France; Service de Chirurgie Orthopédique et Traumatologique, Hôpital Henri Mondor AP-HP, CHU Paris 12, Université Paris-Est, 51 avenue du Maréchal de Lattre de Tassigny, 94000 Créteil, France

* Correspondence: guillaume.haiat@cnrs.fr ;

## Abstract

Open-wedge High-Tibial Osteotomy is the most commonly employed technique addressing varus knee osteoarthritis. During this procedure, the osteotome passes through cortical and trabecular bone. A common complication is lateral hinge fractures, which depends upon when the surgeon stops impacting the osteotome. The aim of the present study is to determine whether an instrumented hammer can be employed to detect i) the bone type (cortical or trabecular) surrounding the osteotome tip and ii) when bone is fractured by the osteotome. Seventeen lower-limb specimens were obtained from nine fresh human anatomical subjects. The osteotome was impacted using the instrumented hammer until rupture. Each impact signal was recorded and analyzed. An algorithm was developed to detect the transitions between cortical and trabecular bone and *vice versa*. Detection was based on the relative variation of $\tau$ derived from the force signal as a function of time as the time difference between the second and first impact peaks. The values of $\tau$ were significantly lower in cortical bone compared to trabecular bone. Considering the arrival of the osteotome tip in trabecular bone during a medial, a difference of less than 2 impacts between the surgeon and the algorithm was obtained for 71 % of the cases. For the lateral osteotomy it was obtained in 81% of the cases. This cadaveric study demonstrated that the instrumented hammer could prevent impacts leading to hinge rupture, which suggests that future clinical trials are warranted.

236 Words

## 1. Introduction

The knee is a complex and essential joint that bears significant loads while enabling motion[1]. The distribution of medial and lateral forces depends on knee alignment and varies according to different weight-bearing tasks[2]. Osteoarthritis and degeneration of the knee joint are the most prevalent joint disease among adults worldwide[3]. It often leads to a varus deformity, which increases load transmission through the already degenerated compartment[4]. When conservative treatments fail, Total Knee Arthroplasty (TKA) is considered as a successful and commonly performed procedure for managing osteoarthritis[5]. For younger patients with moderate yet symptomatic arthritis, High Tibial Osteotomy (HTO) represents a joint-preserving alternative to TKA[6]. Open-Wedge High Tibial Osteotomy (OW-HTO) is most commonly used to correct varus malalignment[7]. This procedure involves an osteotome that is struck with a mallet to cut through bone and cartilage tissue (see Fig. 1) [8].

The most frequent complication is a lateral hinge fracture (LHF), occurring in approximately 15–25% of cases [9,10]. Intra-articular fractures of the lateral tibial plateau are particularly serious due to their extension into the subchondral bone of the compartment receiving the shifted load (see Fig. 1) [11,12]. Despite the clinical relevance of these complications, few solutions exist to assist surgeons during the procedure and help prevent fractures. Gulagaci et al. demonstrated that placing a protective hinge wire intersecting the osteotomy plane can significantly reduce the incidence of perioperative hinge fractures[13]. Jeong et al. developed a machine learning model to predict LHFs[14]. Their study included data from 439 knees with medial osteoarthritis treated using medial OW-HTO. Utilizing both X-ray and Computed Tomography (CT) imaging, they developed two models, one combining Light Gradient Boosting Machine with a multilayer perceptron, and the other combining logistic regression with a multilayer perceptron, both showing high predictive performance for LHF occurrence.

Our team has developed a new device that does not alter the standard surgical workflow and has been used to monitor acetabular cup and femoral stem insertion during total hip arthroplasty[15-20]. This device consists of a hammer equipped with a force sensor, which records the time-varying force signal at the interface between the hammer and the impacted tool (e.g., the osteotome). Each recorded signal corresponds to a single impact and provides information about the mechanical properties of the bone tissue surrounding the osteotome. A preliminary study showed that this instrumented hammer can estimate both the thickness and the mechanical properties (*e.g.:*, Young's modulus) of the material being cut[21].

Further *ex vivo* and anatomical studies demonstrated that the instrumented hammer can detect fractures and transitions between bone types, such as from nasal to frontal bone during rhinoplasty[22,23]. Moreover, it has been shown to predict the rupture of bovine trabecular bone and plywood plates up to 2.6 impacts before failure[24]. More recently, this instrumented hammer successfully detected the rupture of the pterygoid plates *ex vivo*[25]. A preliminary HTO study on porcine tibia also demonstrated that the instrumented hammer could detect transitions from cortical to trabecular bone (and *vice versa*) with an average lead time of 0.8 and 3.6 impacts[26].

However, the use of this instrumented hammer has yet to be validated in a context closer to clinical conditions. Such validation would be crucial for the development of a real-time decision support tool for surgeons during HTO.

The aim of the present study is to determine whether an instrumented hammer can be used to (i) identify the bone type surrounding the tip of the osteotome and (ii) detect when a bone fracture occurs. To do so, the osteotome was impacted with the instrumented hammer as it progressed through the initial cortical layer and into trabecular bone. The osteotomy was completed by continuing the osteotome's path toward the lateral hinge until fracture occurred. Each impact signal was recorded and analyzed.

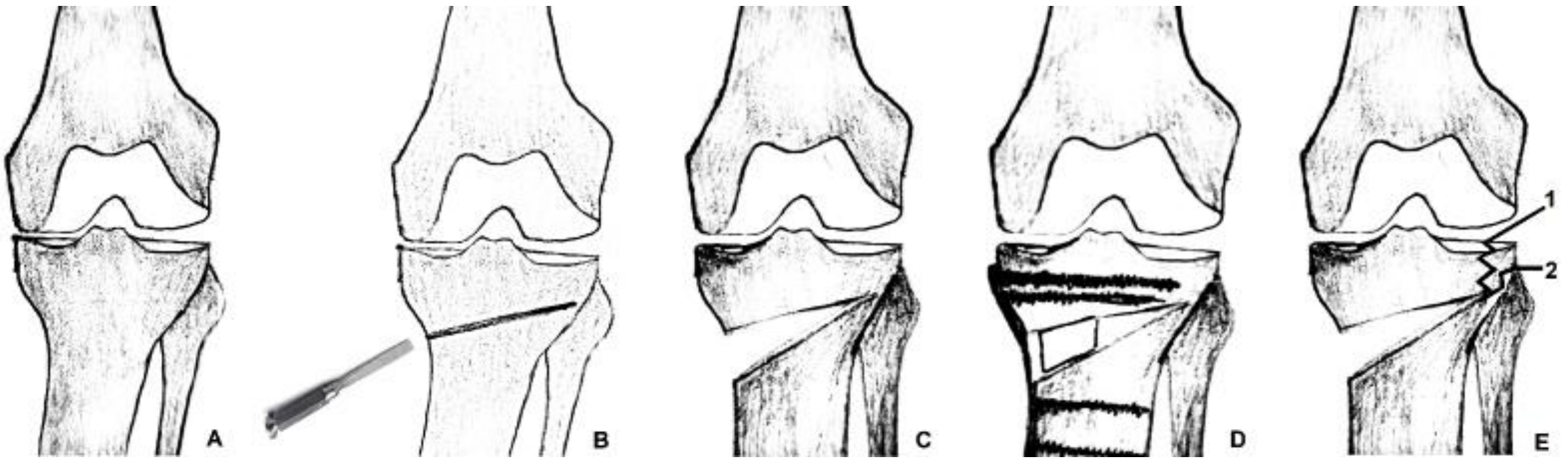

Figure 1: Schematic representation of an open-wedge high tibial osteotomy: A. Varus arthritis; B. Osteotomy with an osteotome; C. Open wedge osteotomy; D. Internal fixation with a metal plate, E. Fractures during open wedge osteotomy. 1. Articular plateau fracture 2. Lateral hinge fracture

## 2. Materials and Methods

### *2.1. Specimens*

Seventeen lower-limb specimens were obtained from nine fresh human cadavers, 5 women and 4 men with a mean age of 78.6 ±2.7 years old years old. The inclusion criteria were: (1) fresh lower extremity specimens from adult donors, and (2) intact lower limbs. Exclusion criteria were: (1) any history of orthopedic surgery on the lower limbs, and (2) pre-existing skeletal deformities of the lower limbs. The study protocol was approved by the ethics committee of the Fer-à-Moulin Surgical School (AP-HP).

### *2.2. Instrumented Hammer*

A 900 g surgical mallet (32-6900-90, Zepf, Tuttlingen, Germany) was used to impact an osteotome (MB681R, HIBBS Retti, B. Braun Aesculap) with a 25 mm-long cutting edge. For each impact, the osteotome was manually held in a manner consistent with standard clinical practice. A dynamic piezoelectric force sensor (200C20, PCB Piezotronics, Depew, NY, USA), with a compression measurement range up to 88.96 kN, was mounted at the center of the mallet's impacting surface to measure the force transmitted to the osteotome during each strike. The force signal was acquired using a data acquisition module (NI 9234, National Instruments, Austin, TX, USA) operating at a sampling rate of 51.2 kHz and a resolution of 24 bits. Each signal was recorded over a 20 ms time window using a LabVIEW interface (National Instruments, Austin, TX, USA) and transferred to a computer for analysis.

### *2.3. Experimental procedure* s

The experiments were conducted by an experienced orthopedic surgeon. Each anatomical specimen was positioned supine. A 30 cm vertical skin incision was made at the midline of the knee, followed by the creation of a lateral soft-tissue flap to expose the lateral aspect of the tibia and fibula, thereby allowing visualization of the osteotome exit point. Muscles and ligaments were carefully dissected to provide direct access to the medial side of the tibia. A reference point was identified 3.5 to 4 cm distal to the medial joint line, and an oblique osteotomy trajectory was defined, extending from this point to a location approximately 1 cm below the lateral articular margin of the tibia, directed toward the tip of the fibular head. The tibial osteotomy was then performed from the medial to the lateral aspect of the tibia. The exit of the osteotome tip was continuously monitored using a video camera (L-920M3, Spedal, Taiwan) throughout the procedure (see Fig. 2).

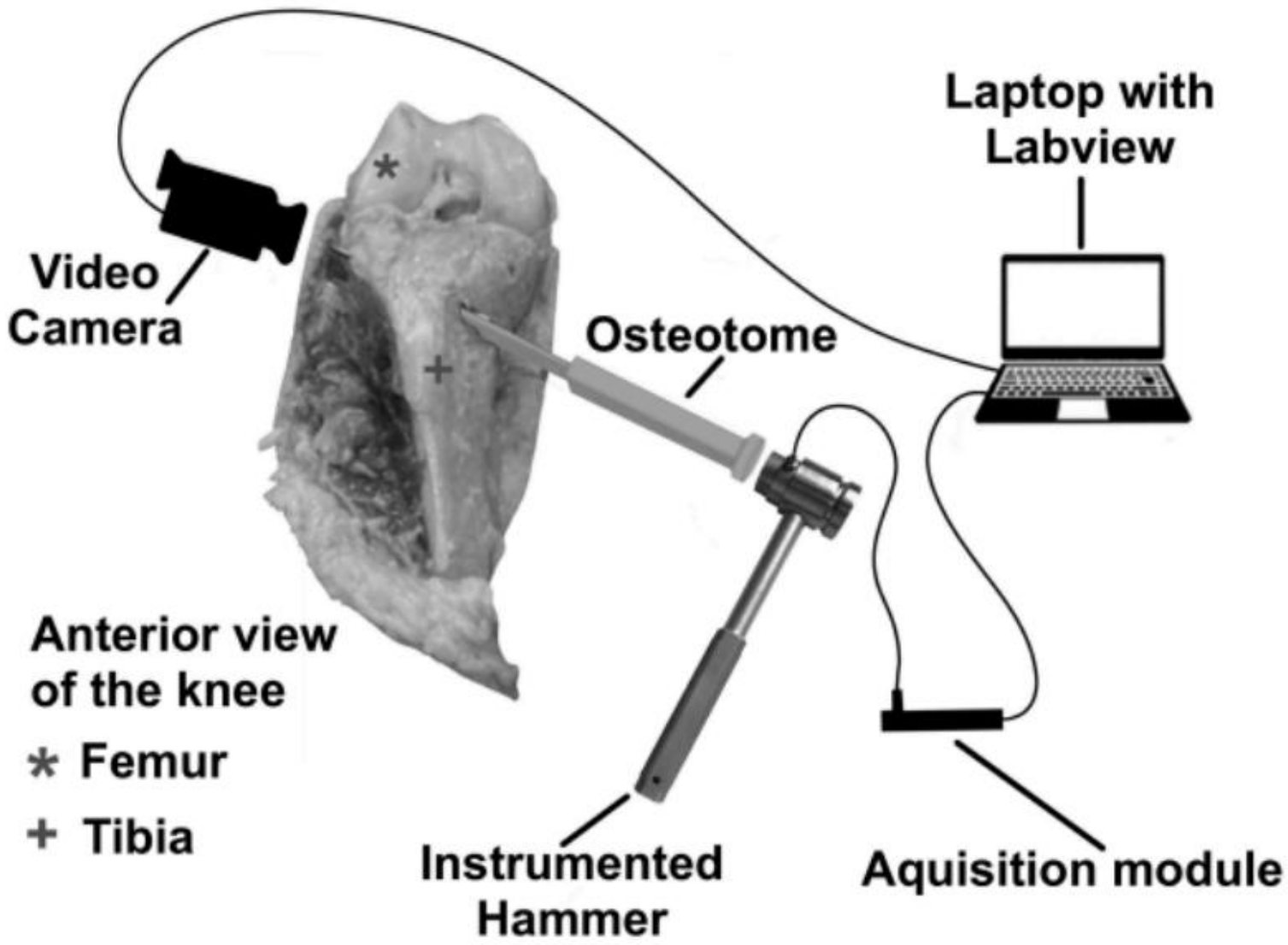

Figure 2. Schematic representation of the experimental configuration showing the anatomical subject lower limb, the osteotome and the hammer instrumented with the piezoelectric sensor. The location of the video camera is also shown.

The objective of the following series of impact experiments was to evaluate whether the instrumented hammer could (i) assess the progression of the osteotome through different types of bone tissue, and (ii) detect the initiation of a lateral hinge fracture. These results were then compared to the surgeon's proprioceptive assessment.

The first series of impacts (medial osteotomy), labelled as #1 in Fig. 3, was performed using impact with a relatively high energy. This phase corresponds to the initial step of HTO, intended to penetrate the medial cortical bone. It ends when the osteotome tip reaches trabecular bone. The peak force $F$ recorded for each impact ranged from 1,000 to 16,000 N. The total number of impacts realized during each medial osteotomy procedure for each bone sample #$j$ with $j \in [1, 17]$ was noted $Q_m(j)$.

The second series of impacts (lateral osteotomy) aimed at obtaining a complete fracture of the cortical hinge by continuing the osteotomy through the lateral aspect of the tibia. Note that hinge fracture should be strictly avoided in a clinical context. This phase was divided into two steps. In the first step (#2.1 in Fig. 3), the osteotome progressed through trabecular bone until the surgeon identified the transition into cortical bone. In the second step (#2.2 in Fig. 3), impacts were continued until the osteotome tip exited the lateral cortical bone. This exit was confirmed *via* video monitoring. The first impact at which the osteotome tip became visible on the video feed, and thus corresponding to bone exit, was denoted $N_e(j)$ for each sample #$j$. The peak force $F$ for these impacts also ranged from 1,000 to 16,000 N. The total number of impacts delivered during the lateral osteotomy for each bone sample *#j* was noted $Q_l(j)$ with $j \in [1, 16]$. One sample was excluded from the analysis due to soft tissue obscuring the video, making observation of the osteotome exit impossible.

During each series of experiments, the surgeon indicated when a transition between bone types (i.e., cortical to trabecular, or *vice versa*) was perceived. The impact number corresponding to i) the surgeon's detection of the osteotome tip entering cortical bone was noted $P_c(j)$, ii) the surgeon's detection of the osteotome tip entering trabecular bone was noted $P_t(j)$ and iii) the surgeon's detection of the osteotome tip exiting cortical bone was noted $P_e(j)$. For four specimens, the osteotome exit was detected by video before the surgeon perceived it, prompting termination of the procedure to avoid further unnecessary impacts. For these samples, $P_e(j)$ was not defined ($P_e$ = *N.A.*).

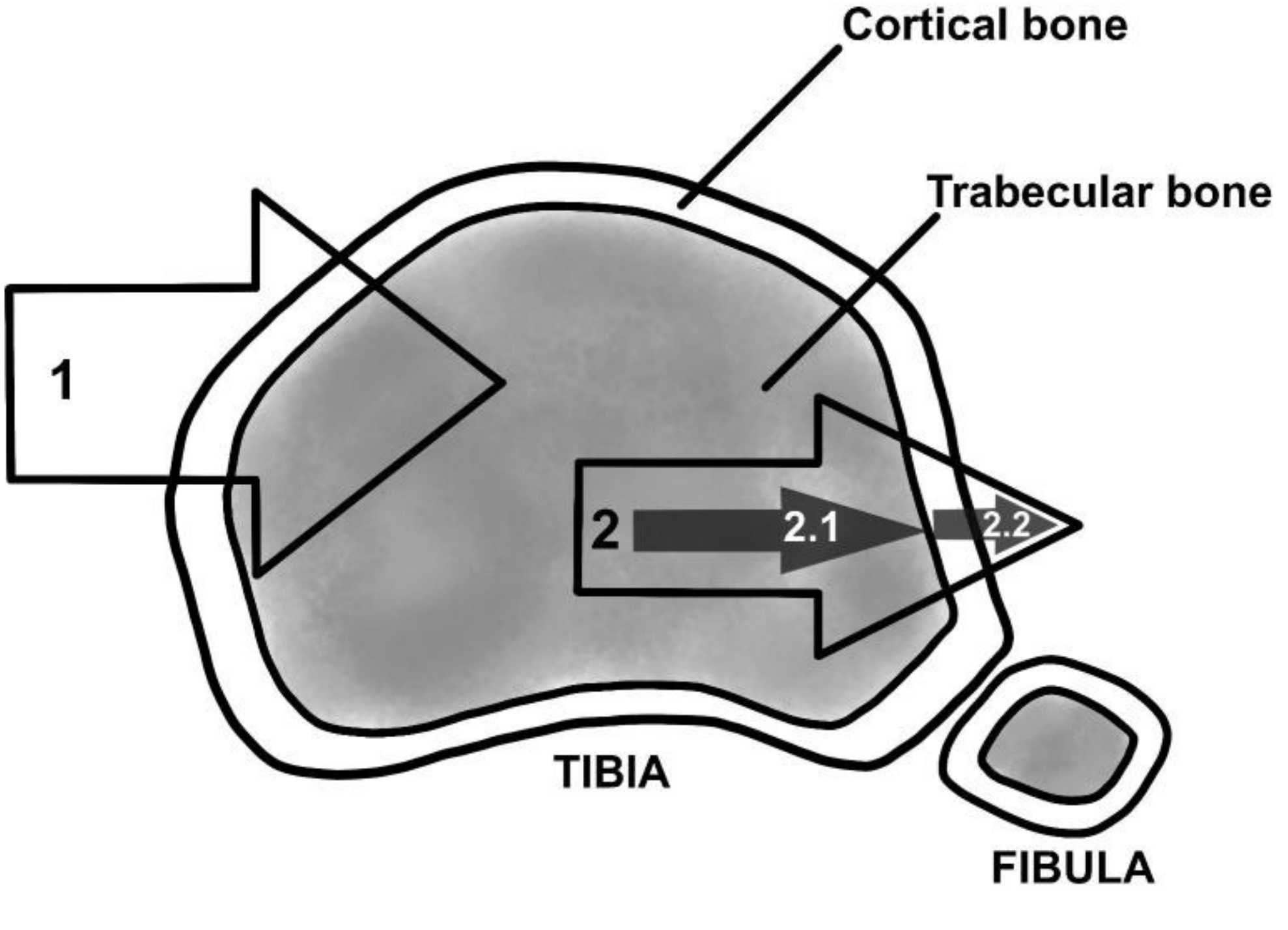

Figure 3. Axial view of the osteotomies. 1. Medial osteotomy. 2. Lateral osteotomy. 2.1. Lateral osteotomy: arrival on the cortical bone. 2.2. Lateral osteotomy: exit of the osteotome

### *2.4. Data processing*

The data analysis was conducted following the same methodology as in Hubert et al[21]. The force sensor described earlier was used to measure the time dependence of the force applied to the osteotome during each impact, resulting in a signal noted *s(t)*. Figure 4 illustrates two examples of typical signals obtained for cortical bone and trabecular bone.

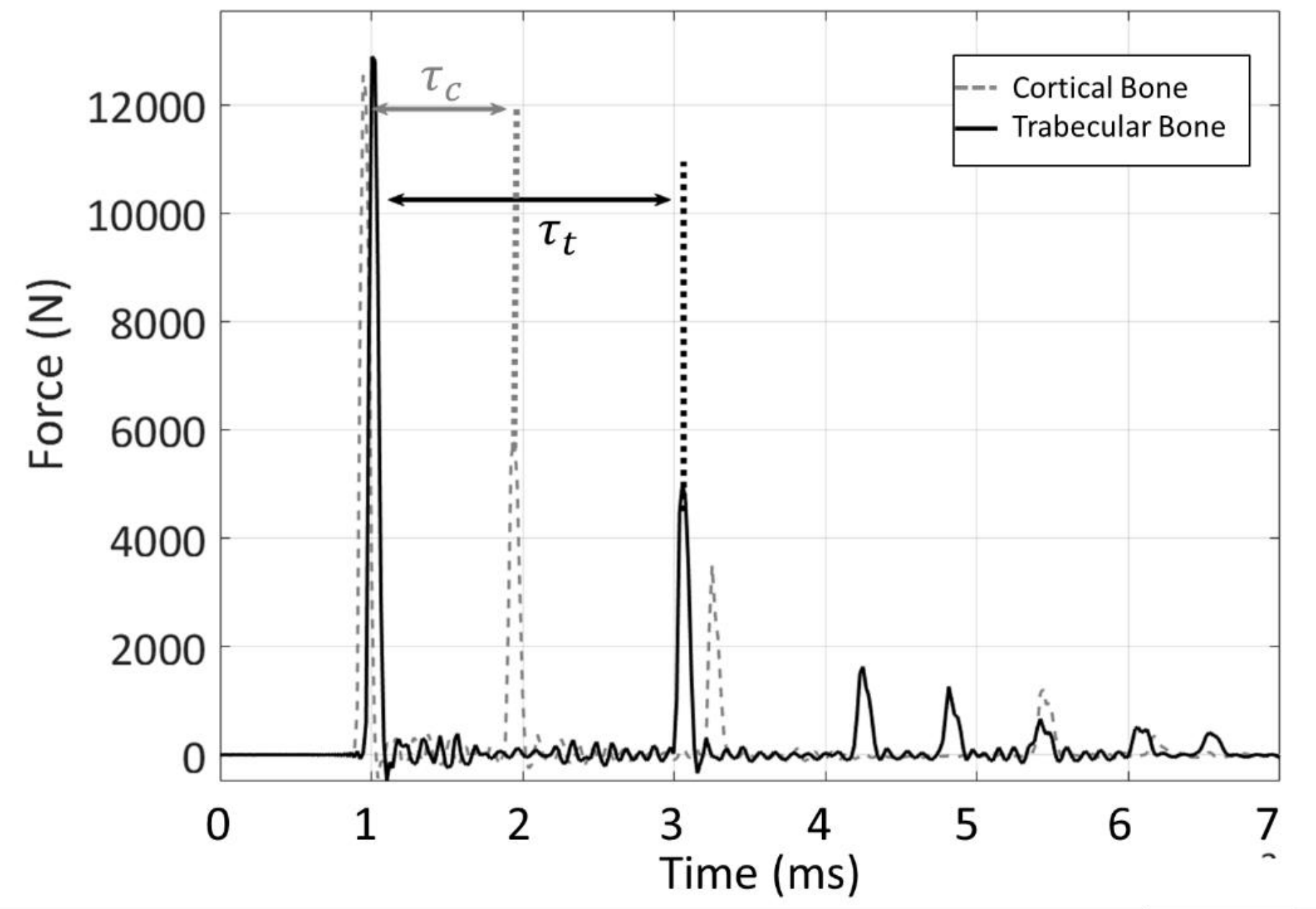

Figure 4: Example of two signals *s(t)* corresponding to the variation of the force as a function of time obtained for a given impact of the instrumented hammer on the osteotome in contact with cortical and trabecular bone.

A dedicated signal processing technique was applied to analyze *s(t)*, using the information obtained from the various peaks present in the signal. As already done in previous studies by our group, an indicator called $\tau$ was determined and corresponds to the difference between the times of the maximum value of the second and first peaks of the signal[21,22,25].

The data were analysed using the method described in the appendix.

## 3. Results

### *3.1. Behavior of the indicator τ as a function of the impact number*

#### *3.1.1. First series: medial osteotomy*

Figure 5 shows the variation of $\tau$ as a function of the impact number for a given medial osteotomy. A progressive increase of the value of $\tau$ is obtained after around 8 impacts, which may correspond to a transition from medial cortical bone to trabecular bone. For the sample corresponding to Fig. 5, the surgeon detects the transition from cortical to trabecular bone for the impact $P_t$ =11.

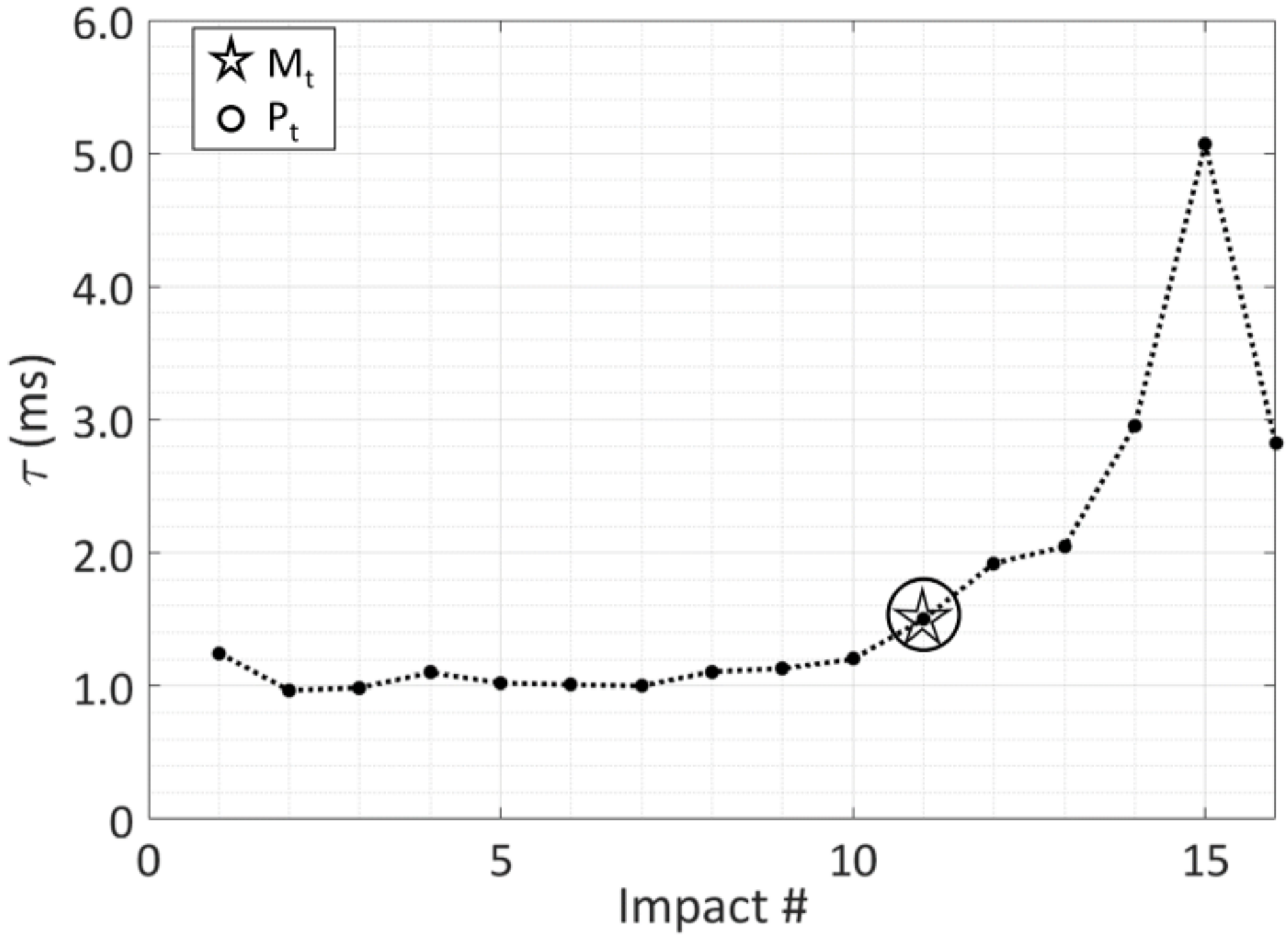

Figure 5. Variation of $\tau$ obtained during the medial osteotomy for a given sample. The star corresponds to the impact number $M_t$ of the transition from the medial cortical bone layer to the trabecular bone predicted by the instrumented hammer with the algorithm after optimization. The circle corresponds to the impact number $P_t$ determined by the surgeon for the transition from the medical cortical bone layer to the trabecular bone.

### *3.1.2. Second series: lateral osteotomy*

Figure 6 shows the variation of $\tau$ as a function of the impact number for a given lateral osteotomy. A decrease of the value of $\tau$ is first obtained until around 9 impacts, which may correspond to a transition from trabecular bone to the lateral cortical bone. Then, the value of $\tau$ is shown to increase until the end of the lateral osteotomy, which may correspond to the exit of the osteotome. For the sample corresponding to Fig. 6, the surgeon detects the transition from trabecular to cortical bone for the impact $P_c$ = 5, and the exit of the osteotome for the impact $P_e$ = 19.

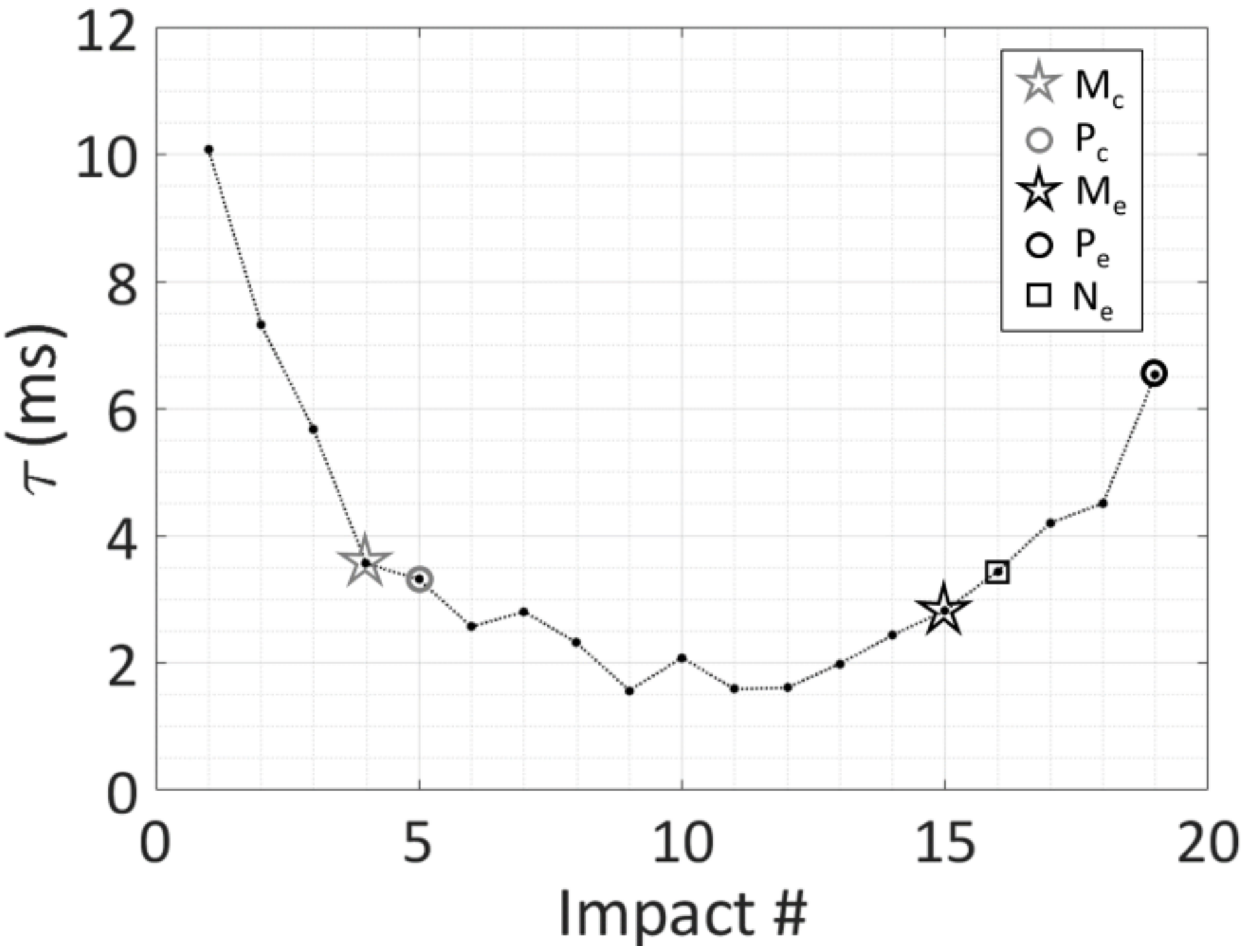

Figure 6. Variation of $\tau$ obtained during the lateral osteotomy for a given sample. The grey star corresponds to the impact number $M_c$ of the transition from the trabecular bone to the lateral cortical bone predicted by the instrumented hammer with the algorithm after optimization. The grey circle corresponds to the impact number $P_c$ determined by the surgeon. The black star corresponds to the impact number $M_e$ of the transition from the trabecular bone to the exit of the osteotome predicted by the instrumented hammer with the algorithm after optimization. The black circle corresponds to the impact number $P_e$ determined by the surgeon. The square corresponds to the impact number $N_e$ predicted by the video camera for the exit of the osteotome.

### *3.2. Optimization procedure*

Figure 7A shows the variation of $\tilde{\Delta}_t$ as a function of $k_t$. Figure 7B the variation of $\tilde{\Delta}_c$ as a function of $k_c$. Figure 7C the variation of $\tilde{\Delta}_e$ as a function of $k_e$ for the medial and lateral osteotomy (see section 2.5). The ranges of variation of $k_t$, $k_e$ and $k_c$ were chosen to emphasize a local minimum but we checked that all values comprised between 1 and 5 (respectively between 0.2 and 1) do not lead to any other local minimum (data not shown). The results led to the following optimized values: $k_t = 1.21$, $k_c = 0.605$ and $k_e = 1.19$.

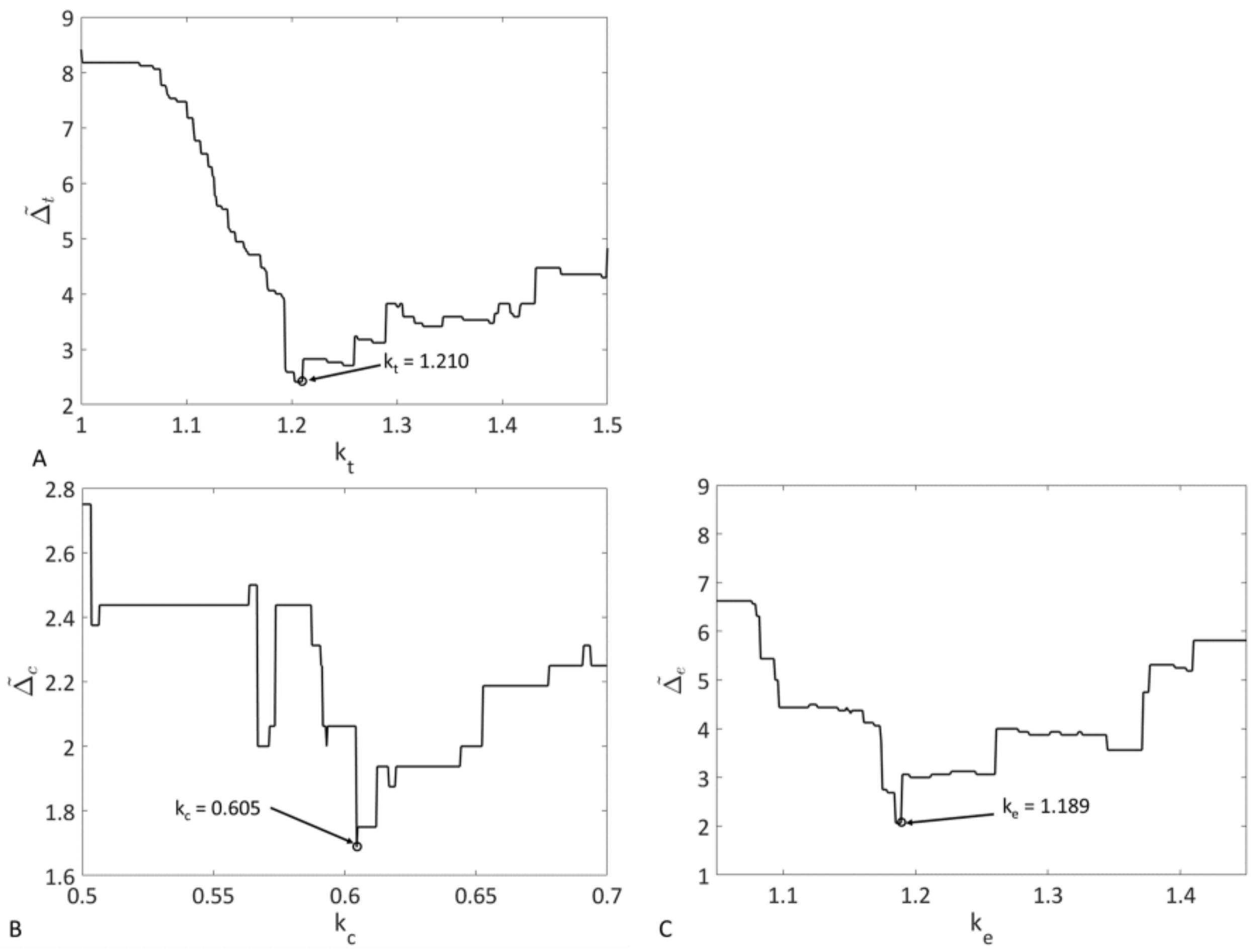

Figure 7: Results used to find the optimal value of $k_t$, $k_c$ and $k_e$ : A. Variation of the discrepancy $\tilde{\Delta}_t$ as a function of $k_t$ ; B. Variation of the discrepancy $\tilde{\Delta}_c$ as a function of $k_c$; C. Variation of the discrepancy $\tilde{\Delta}_e$ as a function of $k_e$.

Figures 5 and 6 show the results obtained with the optimized versions of the three algorithms. For the medial osteotomy, the optimized algorithm detects the transition at the same impact as the surgeon ($M_t = P_t = 11$). For the lateral osteotomy, the optimized algorithm detects the arrival on cortical bone for the impact number $M_c = 4$, whereas the surgeon detects it on the impact $P_c = 5$. For the exit of the osteotome the algorithm detects it one impact before the video camera ($N_e = 16$ and $M_e = 15$) and the surgeon detects it on the impact $P_e = 19$.

### *3.3. Result for all the database*

#### *3.3.1. Medial osteotomy*

Table 1 shows the average and standard deviation values of $\tau$ for all medial osteotomies ($\#j \in [1; 17]$) for the impacts realized in trabecular and cortical bones according to the algorithm and to the surgeon. For each sample $\#j$, the impacts realized in cortical bone according to the algorithm are defined by $i \in [1; M_t(j)[$ and according to the surgeon by $i \in [1; P_t(j)[$. Conversely, the impacts realized in trabecular bone according to the algorithm are defined by $i \in [M_t(j); Q_m(j)]$ and by $i \in [P_t(j); Q_m(j)]$ for the surgeon.

The average values of $\tau$ obtained in trabecular bone, according to the surgeon are significantly higher than the one in cortical bone, which was shown following an ANOVA analysis ($p < 0.001$, $F=137.1$) The same results were found for the algorithm ($F=121.5$).

| **Material surrounding the osteotome tip** | **According to :** | **Impacts belonging to** | **Mean $\tau$ ± SD (ms)** |
|---|---|---|---|
| **Cortical Bone** | Surgeon | $i \in [1; P_t(j)[$ | 2.08 ± 0.77 |
| | Algorithm | $i \in [1; M_t(j)[$ | 1.9 ± 0.69 |
| **Trabecular Bone** | Surgeon | $i \in [P_t(j); Q_m(j)]$ | 4.06 ± 2.11 |
| | Algorithm | $i \in [M_t(j); Q_m(j)]$ | 3.66 ± 1.89 |

Table 1: Averaged and standard deviation values of $\tau$ obtained in cortical and trabecular bone tissue according to the surgeon and to the optimized algorithm during the medial osteotomy.

Figure 8 shows the distribution of the difference between the impact numbers corresponding to the arrival n trabecular bone predicted by the algorithm and by the surgeon. A perfect match between the surgeon and the algorithm is obtained for 18% of the samples and the difference is smaller or equal to - 2 impacts in 71 % of the cases. More importantly, the algorithm always detects the arrival in trabecular bone before the surgeon, with an average error of 2.4 ± 2.6 impact.

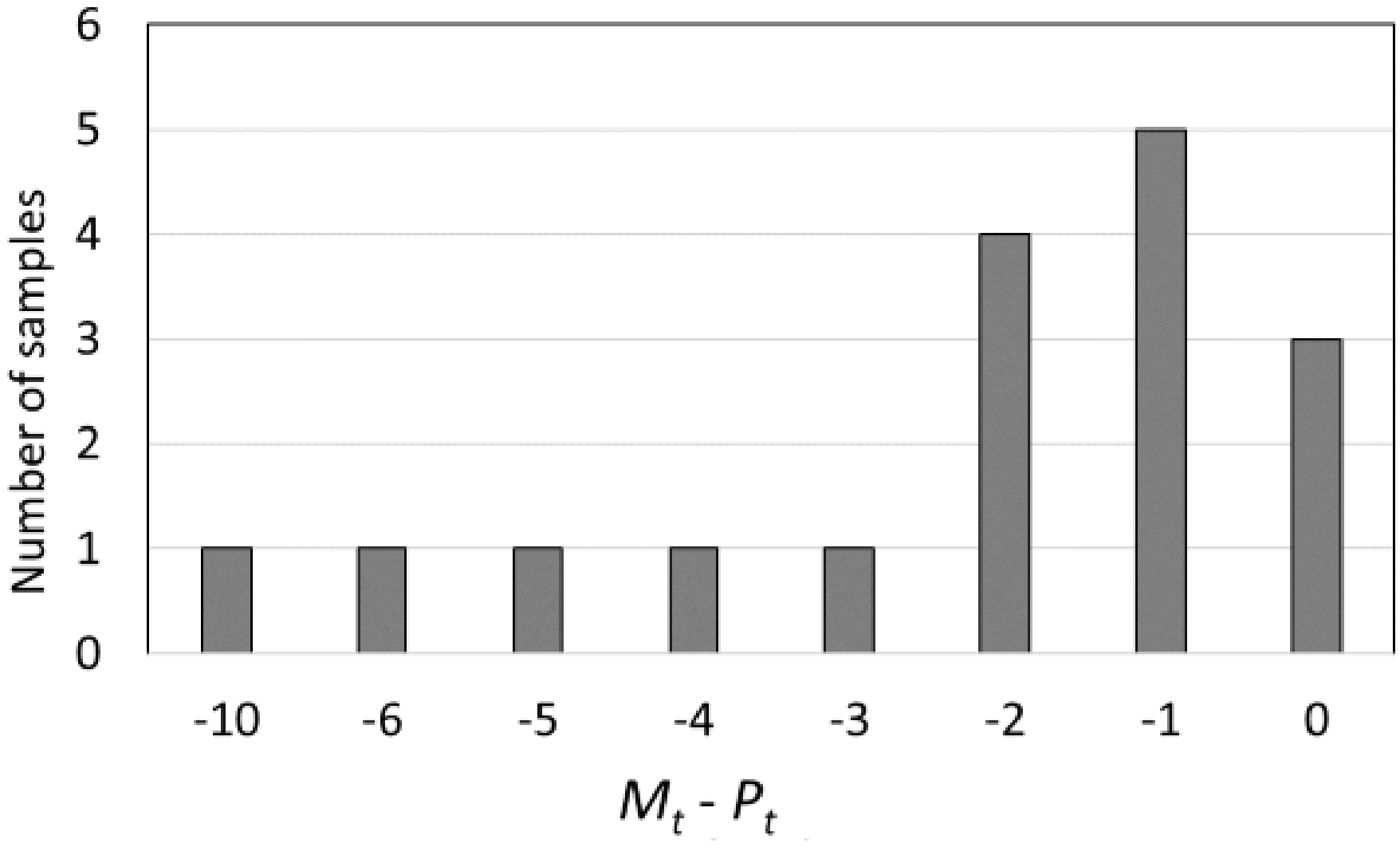

Figure 8. Distribution of the values of $M_t$ - $P_t$, where $M_t$ are the impact number corresponding to the osteotome tip arriving in trabecular bone according to the algorithm, and and $P_t$ to the surgeon proprioception for the medial osteotomy.

### *3.3.2. Lateral osteotomy*

Table 2 shows the average and standard deviation values of $\tau$ for all the lateral osteotomies ($\#j \in [1; 16]$) for the impacts realized in trabecular and cortical bone according to the algorithm and to the surgeon. For each sample $\#j$, the impacts realized in trabecular bone according to the algorithm are defined by $i \in [1; M_c(j)[$ and by $i \in [1; P_c(j)[$ for the surgeon. Conversely, the impacts realized in cortical bone according to the algorithm are defined by $i \in [M_c(j); M_e(j)[$ and by $i \in [P_c(j); P_e(j)[$ for the surgeon.
Table 2 also shows the average and standard deviation values of $\tau$ for the impacts realized after the osteotome tip has exited lateral cortical bone according to the algorithm, to the surgeon and to the video camera. For each sample *#j*, the impacts realized with the osteotome tip in air according to the algorithm are defined by $i \in [M_e(j); Q_l(j)]$, $i \in [P_e(j); Q_l(j)]$ for the surgeon and $i \in [N_e(j); Q_l(j)]$ for the video camera.

The average values of $\tau$ obtained in trabecular bone, in cortical bone and in the air, according to the surgeon are significantly different, which is shown by an ANOVA analysis ($p < 0.001$, F=81.1). The same results are found for the algorithm (F=139.6).

| **Material surrounding the osteotome** | **According to :** | **Impacts belonging to** | **Mean $\tau$ ± SD (ms)** |
|---|---|---|---|
| **Trabecular Bone** | Surgeon | $i \in [1; P_c(j)[$ | 7.17 ± 3.37 |
| | Algorithm | $i \in [1; M_c(j)[$ | 8.34 ± 3.11 |
| **Cortical Bone** | Surgeon | $i \in [P_c(j); P_e(j)[$ | 3.69 ± 1.03 |
| | Algorithm | $i \in [M_c(j); M_e(j)[$ | 3.49 ± 1.14 |
| **Air** | Surgeon | $i \in [P_e(j); Q_l(j)]$ | 6.05 ± 2.88 |
| | Algorithm | $i \in [M_e(j); Q_l(j)]$ | 4.93 ± 1.89 |
| | Video camera | $i \in [N_e(j); Q_l(j)]$ | 5.12 ± 2.03 |

Table 2: Averaged and standard deviation values of $\tau$ obtained with the osteotome tip in trabecular bone, cortical bone and in air according to the surgeon and to the optimized algorithm during the lateral osteotomy. The last line shows the results obtained with the video camera.

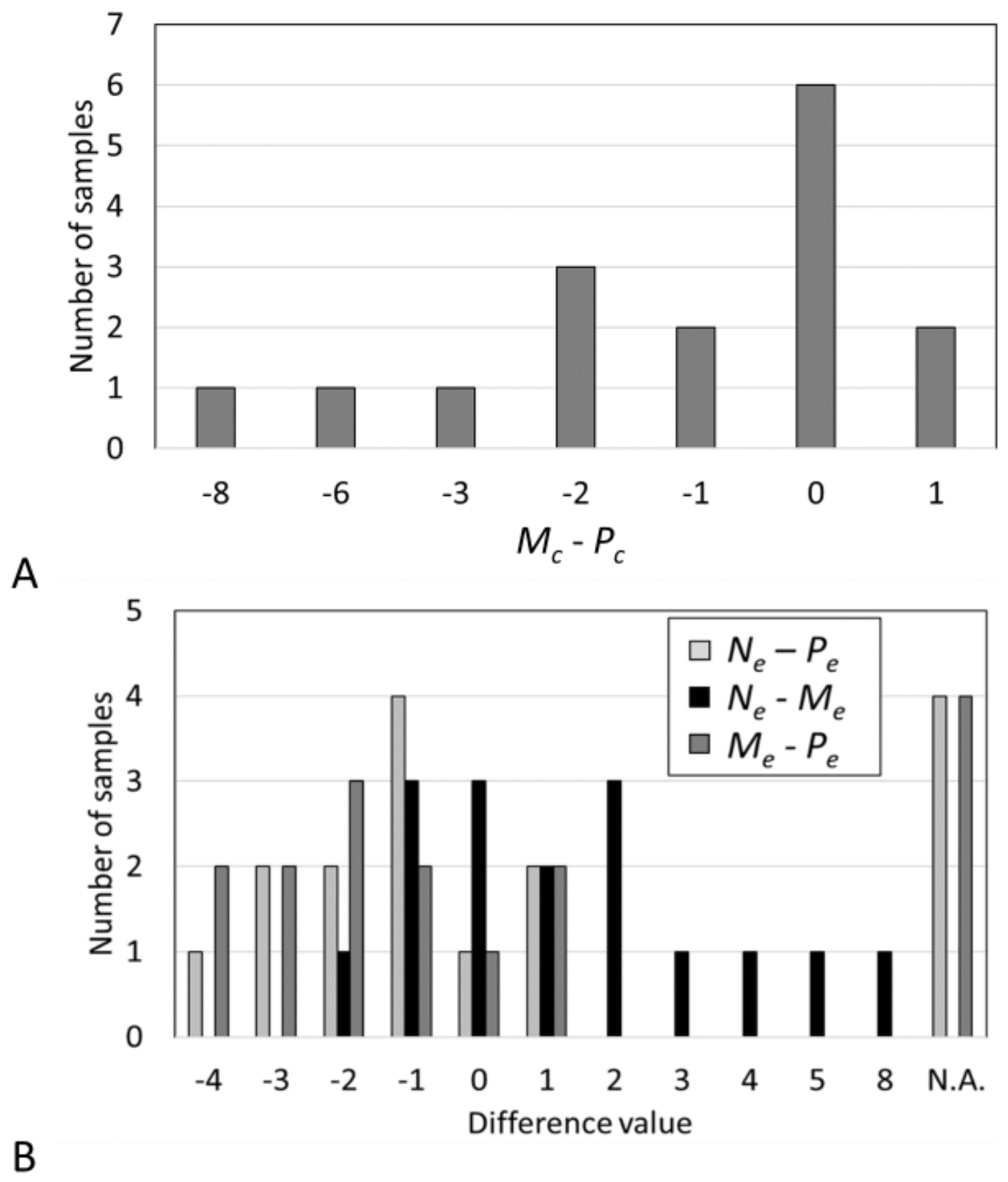

Figure 9. A: Distribution of the values of $M_c - P_c$, where $M_c$ and $P_c$ are the impact number corresponding to the osteotome tip arriving in cortical bone according to the algorithm and $P_c$ to the surgeon proprioception, for the lateral osteotomy. B: Distribution of the values of $N_e - P_e$, $N_e - M_e$, and $M_e - P_e$, where $M_e$ are the impact number corresponding to the osteotome tip exiting cortical bone according to the algorithm, $P_e$ according to the surgeon proprioception and $N_e$ according to the video camera, for the lateral osteotomy. ‘N.A.’ indicates that for 4 samples, the surgeon was unable to determine the impact corresponding to the exit of the osteotome tip.

Figure 9A shows the distribution of the difference between the impact numbers corresponding to the arrival in cortical bone predicted by the algorithm and by the surgeon. A perfect match between the surgeon and the algorithm is obtained for 37% of the samples and the difference is comprised between - 2 and 2 impacts in 81 % of the cases. More importantly, the algorithm always detects the arrival in trabecular bone before the surgeon, except in two cases where the error is equal to one impact only, with an average error of 1.7 ± 2.3 impact.

Figure 9B shows the results corresponding to the exit of the osteotome. For 50 % of the samples, the absolute value of the difference between the impact corresponding to the exit of the osteotome predicted by the video camera and the algorithm is equal or inferior to 1 impact. It is 31% between the algorithm and the surgeon, and 44% between the video camera and the surgeon. The algorithm detected the osteotome exit later than the video camera in 25% of the specimens. In three of these cases, the delay was limited to one impact, and in one case, to two impacts. In contrast, the surgeon identified the exit later than the video camera in 56% of the specimens. Among these, the delay was one impact in four cases, two impacts in two cases, three impacts in two cases, and four impacts in one case. The average difference value between the video camera and the algorithm detection for all the samples is 2 ± 2.1 impacts.

'N.A.' values in Fig. 9B correspond to the four samples for which the surgeon could not detect the exit of the osteotome ($P_e$ = N.A.). Interestingly the two highest values of $N_e - M_e$ (equal to 5 and 8) also correspond to two of these samples where the surgeon could not detect the exit of the osteotome.

## 4.Discussion

The originality of this study was to explore the feasibility of using an instrumented hammer to determine changes of the material surrounding the osteotome tip during HTO in anatomical subjects. The findings demonstrate the potential of this device to potentially become a decision-support system for surgeons performing osteotomies, such as HTO. A criterion based on the variation of $\tau$ was validated to accurately identify the impact number corresponding to the transition between the first cortical bone layer and trabecular bone, as well as between trabecular bone and the second cortical bone layer.

### *4.1. Physical explanation of $\tau$*

Hubert et al. provided a physical interpretation of the behavior of $\tau$ as a function of the mechanical properties of the material surrounding the osteotome tip using a simple analytical model, which could predict signals corresponding to the variation of the force as a function of time during an impact that were similar to the ones shown in Fig. 4 [21]. The time between the first two peaks (denoted $\tau$), is related to the rigidity of the material around the osteotome tip. This relationship comes from the fact that the osteotome bounces back and forth between the hammer and the bone with a characteristic frequency that can be approximated by the resonance frequency of the system, which is in turn related to the mechanical properties of bone tissue located around the osteotome tip. When the osteotome tip is located in trabecular bone, the resonance frequency is thus lower than when it is located in cortical bone, which explains the highest values of $\tau$, as shown in Tables 1 and 2. As demonstrated in Table 2, $\tau$ is significantly higher when the tip of the osteotome is located in air (post-exit) compared to when it is located in trabecular and cortical bone.

### *4.2. Comparison of the performances of the different methods*

As indicated in Figs. 8 and 9, there are several differences in the prediction of the transitions between the two methods, which may be explained by various sources of errors which are described below.

First, possible errors may come from the orthopaedic surgeon because it is an empirical approach. However, it remains difficult to assess quantitatively the error corresponding to the estimation of $P_t$, $P_c$ and $P_e$.

Second, no video analysis was performed herein for the medial impacts (noted #1 in Fig. 3) nor for the first part of lateral impacts (noted #2.1 in Fig. 3). Such analysis was done in our previous animal study by examining bone tissue after it has been cut by the osteotome and using an osteotome with engraved graduations.[26] It was not possible to use the same approach herein because we could not extract bone tissue from the anatomical subjects for ethical reasons, thus preventing an objective and visual comparison for the performance of the instrumented hammer.

Nevertheless, an agreement of ±2 impacts between the mallet and the surgeon was obtained in 71% of the cases for medial impacts and 81% of lateral impacts, which is of the same order of magnitude of the results obtained in the previous animal study (± 2 impacts for 73% and 93% of cases).[26]

Considering the second part of lateral impacts, an objective method using a video camera was employed to monitor the exit of the osteotome. A good agreement was obtained between the video and the instrumented hammer, with a difference of less than 2 impacts for 12 samples. In contrast, discrepancies between the video and the surgeon's proprioception for the osteotome exit were more important. Specifically, the surgeon was late in his prediction in 57% cases and failed to detect the exit for 4 samples. These discrepancies could be attributed to some of the limitations described above such as the age of the anatomical subjects involved or the removal of soft tissues which may have impacted the surgeon's proprioception.

Despite the aforementioned limitations, for 15 out of 17 samples (88%) of the medial osteotomies, $P_t$ - $M_t$ is lower or equal to 5, while for the two other samples, (#11 and #14) the relatively important discrepancy may be explained by the fact that $\tau$ doesn't show a steady increase. Moreover, for these two samples, the value of $\tau(M_t)$ is higher than the average value of $\tau$ for all the impact given in the cortical bone (data not shown).

Considering the lateral osteotomies, $P_t$ - $M_t$ is higher than 5 for only two samples (#2 and #11), which may be explained by an early sharp decrease of $\tau$. Interestingly, the two highest values of $N_e - M_e$ (equal to 5 and 8) also correspond to two samples where the surgeon could not detect the exit of the osteotome. This result could correspond to samples with poor bone quality. It should be noted that although the detection is relatively far from $N_e$, it is still detected by the hammer.

### *4.3. Limitations*

This study has several limitations. First, the anatomical specimens do not reflect the clinical population, as high tibial osteotomy (HTO) is typically performed in younger patients. Although the specimens were obtained from elderly donors, cadaveric bone remains representative of living human bone for biomechanical studies. Moreover, aging does not necessarily imply osteoporosis; several of the donors had preserved bone quality and showed no radiological or macroscopic signs of bone fragility. Therefore, the biomechanical behavior observed in this study can reasonably reflect that of living bone in clinical conditions.

Second, the modeling of the LHFas a simple crossing of the osteotome through the lateral cortical bone represents a simplified scenario compared to the complex nature of actual LHF. In this study, it was not possible to reproduce the exact configuration of an intra-articular fracture of the tibial plateau, which typically occurs in a different plane than the osteotomy. Furthermore, soft tissues were removed around the tibia to allow for visualization of the osteotome exit, which differs from real surgical conditions. Although previous work on acetabular cup stability assessment has shown that the presence of soft tissues does not affect the results, this remains to be confirmed in the context of HTO.[27]

Third, limitations are inherent to the use of the instrumented hammer, as the technique is novel and still under development. Only 17 specimens were analyzed, and further studies are needed with larger sample sizes to strengthen the findings.

Fourth, only a single osteotome type was used. Variations in osteotome geometry, such as blade width, length, or curvature, may influence the signals recorded by the instrumented hammer. Future studies should explore the impact of different osteotome designs. Nonetheless, the straight osteotome remains the standard tool in current HTO procedures.

## 5. Conclusions

This study examines the performance of an instrumented hammer in determining the position of the osteotome tip in the context of HTO perfomed with anatomical subjects, and more specifically its transition between trabecular and cortical bone. Detecting this transition is especially important in HTO, because it indicates when impaction should stop to maintain the integrity of the hinge. Our approach could thus prevent unnecessary impacts and potential complications. This present study serves as a proof of concept and demonstrates the overall feasibility of the approach. Future studies should be conducted in clinical settings, which is necessary before recommending its clinical application.

## Acknowledgement

This project has received funding from the project OrthAncil (ANR-21- CE19-0035-03), from the project OrthoMat (ANR-21-CE17-0004) and from the project ModyBe (ANR-23-CE45-0011-02). This project has received funding from the European Research Council (ERC) under Horizon 2020 (grant agreement # 101062467, project ERC Proof of Concept Impactor). We acknowledge the support of the 'Fer-à-Moulin' surgical school, APHP.

## Statements and Declaration

Ethical considerations: The protocol was approved by the ethical committee of the 'Fer-à-Moulin' surgical school, APHP

Consent to participate: not applicable

Consent for publication: not applicable

Declaration of conflicting interest: The author(s) declared no potential conflicts of interest with respect to the research, authorship, and/or publication of this article.

## Appendix : Data analysis

As it was done in [26], a dedicated algorithm was developed in order to detect the transitions between i) cortical and trabecular bone (series 1: Medial osteotomy), ii) trabecular and cortical bone (series 2.1: Lateral osteotomy, arrival in cortical bone) and iii) cortical bone and the exit of the bone (series 2.2: Lateral osteotomy: osteotome exit) based on the variations of $\tau$ as a function of the impact number. According to the results obtained in [21], an increase of $\tau$ is expected when the osteotome tip reaches a medium with a lower Young's modulus, which corresponds to the transition from cortical to trabecular bone and from the cortical bone to the exit. Similarly, a decrease of $\tau$ is expected when the osteotome tip reaches a medium with a higher Young's modulus, which corresponds to the transition from trabecular to cortical bone.

### *1. Medial osteotomy*

For each sample $\#j \in [1;\ 17]$ and each impact $\#i\ \in [1;\ Q_m\ (j)]$, an algorithm was devised to determine the location of the osteotome tip (*i.e.* in cortical or trabecular bone). The osteotome was considered to reach trabecular bone for the first time at impact $\#M_t(j)$ for which the value of $\tau$ is higher than the product of a coefficient $k_t$ and of the average value of the previous values of $\tau$, which corresponds to:

$$\tau\,(M_t(j)) \;>\; k_t \;* \frac{\sum_{l=1}^{M_t(j)-1} \tau(l)}{M_t(j)-1} \qquad (1)$$

where $k_t$ is a constant to be determined using an optimization algorithm (see below).

### *2. Lateral osteotomy*

A similar algorithm was developed in the case of lateral osteotomy. For each sample #$j \in [1; 16]$ and each impact #$i \in [1; Q_l\,(j)]$, the osteotome tip was considered to reach cortical bone for the first time at impact #$M_c(j)$ respecting:

$$\tau\left(M_c(j)\right) < \; k_c * \frac{\sum_{l=1}^{M_c(j)-1} \tau(l)}{M_c(j)-1} \qquad (2)$$

where $k_c$ is a constant to be determined using an optimization algorithm (see below).

Similarly, a third algorithm was developed to determine when the osteotome exited cortical bone. For each sample # $j \in [1; 16]$ and each impact #$i \in [M_c(j); Q_l\,(j)]$ realized after the algorithm described by Eq. 2 detected that the osteotome entered cortical bone, the osteotome tip was considered to exit cortical bone for the first time at impact #$M_e(j)$ respecting:

$$\tau\,(M_e(j)) \;>\; k_e \;* \frac{\sum_{l=1}^{M_e(j)-1} \tau(l)}{M_e(j)-1}, \qquad (3)$$

where $k_e$ is a constant to be determined using an optimization algorithm described below.

Two detection methods were compared to determine the arrival of the osteotome tip in trabecular bone during the medial osteotomy: i) the results predicted by the first algorithm corresponding to the impact #$M_t(j)$ and ii) the proprioception of the surgeon, which corresponds to impact #$P_t(j)$. This comparison led to the determination of the optimal value of $k_t$, which was performed by minimizing the error between the two aforementioned methods given by $\tilde{\Delta}_t$, where $\tilde{\Delta}_t$ corresponds to the average value of $|P_t(j) - M_t(j)|$ for $j \in [1; 17]$.

Similarly, two detection methods were compared to determine the arrival of the osteotome tip in cortical bone during the lateral osteotomy: i) the results predicted by the second algorithm corresponding to the impact #$M_c(j)$ and ii) the proprioception of the surgeon, which corresponds to impact # $P_c(j)$. This comparison led to the determination of the optimal value of $k_c$, which was performed by minimizing the error between the two aforementioned methods given by $\tilde{\Delta}_c$, where $\tilde{\Delta}_c$ corresponds to the average value of $|P_c(j) - M_c(j)|$ for $j \in [1; 16]$.

For the exit of the osteotome, three methods were compared: i) the results predicted by the third algorithm corresponding to the impact #$M_e(j)$, ii) the proprioception of the surgeon, which corresponds to impact #$P_e(j)$and iii) the video camera, which corresponds to impact # $N_e(j)$. The comparison between the video camera and the results obtained with the third algorithm led to the determination of the optimal values of $k_e$, which was performed by minimizing the error given by $\tilde{\Delta}_e$, where $\tilde{\Delta}_e$ corresponds to the average value of $|N_e(j) - M_e(j)|$ for #$j \in [1; 16]$.